%% file: bps.tex
\documentclass[12pt]{article}

\begin{document}

\newcommand{\Mvariable}[1]{{#1}}
\newcommand{\overbar}[1]{{\bar{#1}}}
\newcommand{\imag}{i}

\input{macros}

\begin{titlepage}
\rightline{}

\rightline{hep-th/0405103}

\vskip 2cm
\begin{center}
\Large{{\bf On the polarization of closed strings\\ by Ramond-Ramond fluxes}}
\end{center}

\vskip 2cm
\begin{center}
Vatche Sahakian\footnote{\texttt{sahakian@hmc.edu}}
\end{center}
\vskip 12pt
\centerline{\sl Keck Laboratory}
\centerline{\sl Harvey Mudd}
\centerline{\sl Claremont, CA 91711, USA}

\vskip 2cm

\begin{abstract}
In the Green-Schwarz formalism, the closed string worldsheet of the IIB theory couples to Ramond-Ramond (RR) fluxes through spinor bilinears. We study the effect of such fluxes by analyzing the supersymmetry transformation of the worldsheet in general backgrounds. We show that, in the presence RR fields, the closed string can get `polarized', as the spinors
acquire non-zero vevs in directions correlating with the orientation of close-by D-branes. Reversing the argument, this may allow for worldsheet configurations - with non-trivial spinor structure - that source RR moments.

\end{abstract}

\end{titlepage}
\newpage
\setcounter{page}{1}

\section{Introduction}
\label{intro}

One of the most interesting attributes of string theory is a natural interplay between open and closed string degrees of freedom. For those interested in understanding confining gauge theories or quantum gravity, this duality often translates into a technically useful relation between Yang-Mills theories and gravitational dynamics~\cite{MALDA1,KLEB,WITHOLO}. More formally, the duality involves an intriguing case of a hierarchy between solitonic and perturbative
degrees of freedom in a given theory, as different variables describing the same physics. 

In the program of formulating open/closed string duality in progressively more transparent frameworks, an important task involves understanding the coupling of Ramond-Ramond (RR) fluxes - cast about by
D-branes - to the closed string worldsheet degrees of freedom~\cite{BVW}-\cite{PEDRO}. In this work, our goal is to show that, in the presence of RR fluxes, one may find the spinor degrees of freedom of the worldsheet acquire non-zero vevs. In effect, these fluxes can polarize the worldsheet vacuum 
into non-trivial supersymmetric configurations correlating with the orientation of nearby D-branes.

While rather novel in string theory, the shifting and veving of spinors in the background of solitons is not a surprising phenomenon (see for example~\cite{RUBAKOV}). What makes this case particularly interesting is the fact that it is associated with  
a degeneracy landscape that can potentially yield to interesting worldsheet
`solitons' that source higher order RR moments. This would then necessarily entail an interesting dynamical parameterization of string
theory degrees of freedom - one involving both perturbative and non-perturbative variables.

In this note, we focus on the light-cone Green-Schwarz formalism of IIB closed string theory. We start by developing the relevant
symmetry transformations on the worldsheet, from superspace to component form. 
In particular, we observe that,
in the presence of background fields, the supersymmetry transformation involves a piece that
Lorentz rotates the spinors on the worldsheet so as to preserve the supergravity gauge conditions of
the background. We are then left with three transformation rules to consider: a spinor translation, a Lorentz rotation, and kappa symmetry. 
To preserve the worldsheet light-cone gauge, we need to combine these three symmetry transformations
in a careful recipe. Doing so leaves one with a set of interesting first order (BPS-like) differential equations in
the worldsheet fields that prescribe conditions to be satisfied so that the worldsheet vacuum does not break all the supersymmetries of the background. These equations are found to have an interesting
structure: when RR fluxes are turned on, there is room for supersymmetric vacua with non-trivial vevs for the spinors. We analyze an explicit example and indeed find the non-zero vevs that polarize the worldsheet. We also find flat directions that correlate with the orientation of the D-branes
sourcing the fluxes. And we note that this phenomenon is not manifested with NSNS fluxes.

In Section 2, we present the superspace formalism that is our natural starting point. In Section 3, we describe the simplifications resulting from fixing the light-cone gauge. In particular, we impose a series of conditions on the background fields so as to make the problem tractable. Section 4 lays out
the relevant symmetry transformations of the worldsheet in component form. Section 5 combines
these transformations and formulates the supersymmetry transformation that preserves the light-cone gauge. Section 6 presents the BPS condition in convenient notation. And Section 7 analyzes a particular example that involves worldsheet polarization by an RR flux.

\section{Worldsheet superspace formalism}

We start from the action for the IIB closed string in superspace~\cite{GHMNT}
\bb\label{action}
I=\int d\tau d\sigma \sqrt{-h} h^{ij} \omega V^a_i V^b_j \eta_{ab}+\frac{1}{2} \varepsilon^{ij} V^B_i V^A_j \BB_{AB}\ .,
\ee
where
\bb
V_i^A\equiv \del_i z^M E_M^A\ ;
\ee
and $\BB_{AB}$ is a tensor superfield whose $\theta^0$ component is the Neveu-Schwarz B-field of the IIB theory. The supervielbein is denoted by $E^A_M$, where $A$ is a supertangent-space index and
$M$ is a superspacetime index. In particular, $A$ runs over 10 bosonic polarizations that we 
denote by $a,b,\ldots=0..9$; 16 fermionic polarizations of the Weyl spinor $z^\alpha\equiv \theta^\alpha$; and another 16 polarizations from the complex conjugate $z^\overbar{\alpha}\equiv \theta^\overbar{\alpha}\equiv\overbar{\theta}^\alpha$. Hence, in this notation, the two Majorana-Weyl spinors of the IIB theory are combined into one complex Weyl spinor $\theta$. And the degrees of freedom on the worldsheet are $x^a(\tau,\sigma)$ and $\theta(\tau,\sigma)$. More about the  conventions we use may be found in Appendix A.

In the light-cone gauge, the component form of~\pref{action} truncates to quartic order in $\theta$ for a large class of background field configurations. This form of the action, including the quartic terms, was derived in~\cite{VVSRR} using the normal coordinate expansion technique in superspace. In this paper, we focus on the component form of the supersymmetry transformation so as to formulate more tractable first order differential equations for worldsheet vacua in the presence of background fluxes.

We will write the various symmetry transformations in superspace notation by prescribing a variation
\bb\label{normal}
\delta E^A\equiv \delta z^M E_M^A\ .
\ee
The method of normal coordinate expansion in superspace may be used once again to write the component form of this expression. We will do so in Section~4.1 for future reference; yet, this detailed
form will not be needed in this work. 

We consider three different transformations of the worldsheet degrees of freedom: first, a translation of the spinors by a supervector $\varepsilon^A$ - the backbone of the supersymmetry transformation
\bb\label{eps}
\delta_\varepsilon E^A=\varepsilon^A\ .
\ee
We write $\varepsilon^A$ explicitly in Section~4.2. Second, a supersymmetry transformation
in arbitrary background fields also needs a 
Lorentz rotation
\bb\label{Leq}
\delta_L E^A=E^B \hat{L}_B^A\ .
\ee
We will derive the form of $\hat{L}_B^A$ in Section~4.2 as well. Together, $\delta_\varepsilon+\delta_L$
shall define our supersymmetry transformation. Finally, the
kappa symmetry is prescribed by a supervector $\kappa^A$
\bb\label{kappa}
\delta_\kappa E^A=\kappa^A
\ee
which we will fix in Section~4.3.

\section{Light-cone gauge choice and truncation}

A great deal of simplification is achieved in the task of unraveling superspace into its component form if one is to impose the light-cone gauge conditions
\bb\label{lcgauge}
x^+=p^+\tau\ \ \ ,\ \ \ 
\sigma^+ \theta=0\ .
\ee
Here we have defined
\bb
x^\pm\equiv \frac{1}{2}\lk(x^0\pm x^1\re)\ \ \ ,\ \ \ \sigma^\pm\equiv \frac{1}{2}\lk(\sigma^0\pm\sigma^1\re)\ ;
\ee
and the $\sigma^a$'s are $16\times 16$ gamma matrices
\bb
\lk\{\sigma^a,\sigma^b\re\}=2\eta^{ab}\ .
\ee
For more details about the matrix representation we use, as well as helpful Fierz identities, the
reader is referred to Appendix A and ~\cite{HOWEWEST}.

The gauge condition~\pref{lcgauge} is particularly well adapted for certain background fields configurations. For such a suitable
class of background fields, the worldsheet theory truncates to quartic order in the
$\theta$'s; while the supersymmetry variation truncates to {\em quadratic} order. Henceforth, we
focus exclusively on such backgrounds, and these must obey the following conditions~\cite{VVSRR}:
\begin{itemize}
\item All fields are independent of the $x^+$ and $x^-$ coordinates.

\item All fields carry `-' and `+' indices in pairs (or none at all); for example, a field strength would
have non-zero components $F^{-+i}$ or $F^{ijk}$ but never something like $F^{-ij}$, where $i,j,k,\ldots$ are space directions transverse to the light cone directions `+' and `-'.

\item The background metric is diagonal.

\item All fermionic background fields have zero vevs. In particular, we have no condensates of the gravitino and gaugino.
\end{itemize}
We note that such field configurations include most D-brane geometries if the light-cone directions `+' and `-'  are aligned parallel to the worldvolume of the D-branes.

\section{SUSY, Lorentz rotations, and Kappa symmetry}

In this section, we present the explicit forms of the three transformations~\pref{eps}, \pref{Leq} and~\pref{kappa}. These appear in Sections 4.2 and 4.3. The component form of equation~\pref{normal} on the left hand side of~\pref{eps}, \pref{Leq} and~\pref{kappa} is presented in Section 4.1.

\subsection{Component form of variations}

Equation~\pref{normal} involves variations of all of the worldsheet fields
\bb\label{expan}
\delta E^A=\delta z^a E_a^A+\delta z^\alpha E_\alpha^A-\delta z^\overbar{\alpha} E_\overbar{\alpha}^A\ .
\ee
Hence, we need to unravel the components of the supervielbein. The method of normal coordinate expansion in superspace is well-suited for this task. Appendix B reviews the technique; the reader is also referred to~\cite{GRISARU, ATICKDHAR,VVSRR}. 

First, let us assume that we have arranged for
\bb
\sigma^+\delta\theta=0\ \ \ ,\ \ \ \sigma^+\theta=0\ \ \ \ ,\ \ \ \delta x^m e_m^+=0\ .
\ee
The last two statements follow from the light-cone gauge. The first statement is not necessarily
satisfied in general. However, for all instances where the expansion~\pref{expan} will become useful, we will
see that one indeed has $\sigma^+\delta\theta=0$. These three statements - along with the conditions
on the background fields outlined in Section~3 - severely restrict the structural form of~\pref{expan}.
It is straightforward to show that we necessarily have
\bb\label{s1}
\delta E^+=\delta x^m e_m^+\ .
\ee
\bb\label{s2}
\delta E^-=\delta x^m e_m^--\frac{i}{2} \theta \sigma^- \delta \overbar{\theta}-\frac{i}{2} \overbar{\theta} \sigma^- \delta\theta+\delta x^m e_m^a \Theta_a\ .
\ee
\bb\label{s3}
\delta E^i=\delta x^m e_m^i\ .
\ee
\bb\label{s4}
\delta E^\alpha=\delta \theta^\alpha +\delta x^m e_m^a \Xi_a^\alpha\ .
\ee
\bb\label{s5}
\delta E^\overbar{\alpha}=\delta \theta^\overbar{\alpha} +\delta x^m e_m^a \Xi_a^\overbar{\alpha}\ .
\ee
Indices $i,j,k,\ldots$ label the eight space directions transverse to the light-cone. In these
expressions, the important point is that
$\Xi$ is linear in $\theta$, while 
$\Theta$ is quadratic. All higher order terms cancel because of the light-cone condition and
the form of the background fields: The conditions on the background fields
imply that one cannot absorb a light-cone index `+' or `-' into a field strength; the `+'s and
`-'s should eventually contract with fermion bilinears. But the light-cone gauge and
Fermi statistics for the spinors allow only 
the following non-zero bilinears
\bb
\theta \sigma^{-ij}\theta\ \ \ ,\ \ \ 
\overbar{\theta} \sigma^{-ij} \overbar{\theta}\ \ \ ,\ \ \ 
\theta\sigma^-\overbar{\theta}\ \ \ ,\ \ \ 
\theta\sigma^{-ij}\overbar{\theta}\ .
\ee
It is now easy to see how one arrives at the structural form depicted in the equations~\pref{s1}-\pref{s5}. More explicitely, using the normal coordinate expansion technique outlined in Appendix B, one finds
\bbb\label{c1}
\delta E^a&=&{{{\delta x}}^m}\,e_m^a - \frac{i}{2}\,\theta \sigma^a\delta\overbar{\theta} - 
  \frac{i}{2}\,\overbar{\theta} \sigma^a \delta \theta \nonumber \\
  &-&   \frac{1}{384}{G_{fbcde}\,{{\delta x}^m}\,e_m^f\,\overbar{\theta} \sigma^a \sigma^{bcde}\theta} - 
  \frac{1}{384}{G_{fbcde}\,{{\delta x}_m}\,e_m^f\,\overbar{\theta} \sigma^{bcde} \sigma^a \theta}  \nonumber \\
  &+& \frac{3\, i }{32}\,F_{dbc}\,{{\delta x}_m}\,e_m^d\,\overbar{\theta} \sigma^{bc} \sigma^a\overbar{\theta}
  +   \frac{3\,i}{32}\,{{{\delta x}}_m}\,e_m^d\,\theta \sigma^{bc} \sigma^a \theta\,\overbar{F}_{dbc} \nonumber \\
  &+&   \frac{i}{96}\,F_{bcd}\,{{\delta x}_m}\,e_m^e\,\overbar{\theta} \sigma^{ebcd} \sigma^a\overbar{\theta}+ 
  \frac{i}{96}\,{{{\delta x}}_m}\,e_m^e\,\theta \sigma^{ebcd} \sigma^a \theta \,\overbar{F}_{bcd} \ ;
\eee
\bbb\label{c2}
\delta E^\alpha&=&{{{\delta \theta }}^{\alpha }} - \frac{i}{192}\,G_{abcde}\,{{{\delta x}}^m}\,
   {{\left( \theta \sigma^{bcde} \right) }^{\alpha }}\,e_m^a \nonumber \\
   &+& \frac{3}{16}{\,F_{abc}\,{{{\delta x}}_m}\,{{\left( \overbar{\theta} \sigma^{bc} \right) }^{\alpha }}\,e_m^a}
   +\frac{1}{48} {F_{bcd}\,{{{\delta x}}^m}\,{{\left( \overbar{\theta} \sigma^{abcd} \right) }^{\alpha }}\,
     e_m^a} - {{{\delta x}}^m}\,{{\theta }^{\beta }}\,e_m^b\,\Omega_{b,\beta\alpha}\ ;
 \eee
 \bb\label{c3}
 \delta E^\overbar{\alpha}=\overline{\delta E^\alpha}\ .
 \ee
In these expressions, the indices $a,b,c,\ldots$ run over all ten spacetime directions. And we are using the standard notation for the antisymmetrized gamma matrix basis; \ie\ $\sigma^{ab\ldots}\equiv \sigma^{[a}\sigma^{b\ldots]}$. Let us also identify the various IIB bosonic fields appearing in~\pref{c1}-\pref{c3}:
\begin{itemize}
\item $F_{abc}$ includes the NSNS and RR 3-form fluxes; writing $F_{abc}=F^R_{abc}+i F^I_{abc}$, we have
\bb\label{chi}
F_{abc}^R=\frac{1}{2} e^{-\phi/2} H^{(1)}_{abc}\ \ \ ,\ \ \ 
F^I_{abc}=\frac{1}{2} e^{\phi/2}\lk(\chi H^{(1)}_{abc}+H^{(2)}_{abc}\re)\ .
\ee
where $H^{(1)}$ and $H^{(2)}$ are respectively the NSNS and RR fields, $\phi$ is the dilaton, and $\chi$ is the RR axion.

\item $G_{abcde}$ is the NSNS 5-form field strength.

\item And $\Omega_{a,\alpha\beta}$ is the spacetime connection.
\end{itemize}

The details in~\pref{c1}-\pref{c3} will not be important to the upcoming analysis.

\subsection{Supersymmetry transformation}

As mentioned earlier, the supersymmetry transformation involves both~\pref{eps} and~\pref{Leq}. In general, we have
\bb\label{trans1}
\delta_{\varepsilon+L} E^A_M=\del_M \varepsilon^A+\varepsilon^C \hat{\Omega}_{MC}^{\ \ \ \ A}+\varepsilon^B \hat{T}_{BM}^{\ \ \ A}+E^B_M \hat{L}_B^A\ .
\ee
where $\hat{T}_{BM}^{\ \ \ A}$ is the supertorsion, and $\hat{\Omega}_{MC}^{\ \ \ \ A}$ is the
superconnection. The various components of the supertorsion and superconnection have been
computed in~\cite{HOWEWEST}. We note that, using the conventions of~\cite{HOWEWEST}, the
superconnection includes a U(1) piece under which the $\theta$'s are charged. Similarly, the
rotation matrix $\hat{L}_B^A$ includes a phase rotations under this U(1). We use these
`hat'-ed expressions for connection and rotation to make it easier to
compare to existing conventions in the literature. We will however undo this unnecessary U(1) rotation at the end.

We now want to compute $\varepsilon^A$ and $\hat{L}_B^A$ in background field configurations
conforming to our conditions of Section~3. First, $\varepsilon^A$  should be such that the $\theta^0$
component of the supervielbein is preserved (\ie\ so as to remain within the supergravity 
gauge that has been implicitly fixed); this means we need
\bb
\lk.\delta_{\varepsilon+L} E^A_M\re|_0=0\ .
\ee
Using~\pref{trans1}, we get a transformation with respect to a spinor translation parameter $\varepsilon^\alpha$
and a U(1) phase rotation $q$ 
\bb\label{q1}
\delta_{\varepsilon} E^\alpha=\varepsilon^\alpha-i q\, \theta^\alpha\ .
\ee
\bb\label{q2}
\delta_{\varepsilon} E^\overbar{\alpha}=\varepsilon^\overbar{\alpha}+i q\, \theta^\overbar{\alpha}\ .
\ee
\bb\label{q3}
\delta_{\varepsilon} E^a=\varepsilon^a=i\overbar{\varepsilon}\sigma^a \theta+i\varepsilon\sigma^a \overbar{\theta}\ .
\ee
We note in particular the trivial contributions from the Lorentz rotation 
\bb
\lk.\hat{L}_a^b\re|_0=0\ \ \ ,\ \ \ \lk.\hat{L}_\alpha^\beta\re|_0=i\delta_\alpha^\beta q
\ \ \ ,\ \ \ \lk.\hat{L}_\overbar{\alpha}^\overbar{\beta}\re|_0=i\delta_\overbar{\alpha}^\overbar{\beta} q
\ee
We now
rotate away the phase $q$ for convenience to conform to more conventional worldsheet supersymmetry transformation rules since this phase is not needed to preserve the supergravity gauge. Hence, we set $q\rightarrow 0$ in~\pref{q1} and~\pref{q2}

In general background fields, the transformation $\delta_\varepsilon$ also upsets
the superconnection. To preserve the $\theta^0$ components
of the superconnection, one needs to add a Lorentz rotation
\bb\label{sugra2}
\lk.\delta_{\varepsilon+L}\hat{\Omega}_{MA}^{\ \ \ \ B}\re|_0=0\ .
\ee
And we know
\bb
\delta_{\varepsilon+L}\hat{\Omega}_{MA}^{\ \ \ \ B}=-\del_M \hat{L}_A^B+\varepsilon^N \hat{R}_{NMA}^{\ \ \ \ \ \ B}\ ,
\ee
where $\hat{R}_{NMA}^{\ \ \ \ \ \ B}$ is the superriemann tensor (with a U(1) piece which will not contribute).
It is now easy to find the component form of $\hat{L}_{A}^{\ \ B}$. We first write
\bb\label{rot1}
L_\alpha^\beta=\frac{1}{4} \lk(\sigma^{ab}\re)_\alpha^\beta L_{ab}=\overbar{L}_\alpha^\beta\ .
\ee
\bb
\hat{L}_{ab}=L_{ab}\ \ \ ,\ \ \ \hat{L}_\alpha^\beta=L_\alpha^\beta+i q\delta_\alpha^\beta\ \ \ ,\ \ \ \hat{L}_\overbar{\alpha}^\overbar{\beta}=-L_\alpha^\beta+i q\delta_\alpha^\beta\ .
\ee
From~\pref{sugra2}, one then gets
\bb
L_{ab}=\theta^\alpha \varepsilon^\beta R_{\beta\alpha a b}-\theta^\alpha \varepsilon^\overbar{\beta} R_{\overbar{\beta}\alpha a b}-\theta^\overbar{\alpha} \varepsilon^\beta R_{\beta\overbar{\alpha} a b}+\theta^\overbar{\alpha} \varepsilon^\overbar{\beta} R_{\overbar{\beta}\overbar{\alpha} a b}\ .
\ee
where the relevant components of the superRiemann tensor may be found in~\cite{HOWEWEST}. In terms of the more familiar fields, this is
\bb\label{rot2}
L_{ab}=-\frac{3}{4} i \varepsilon \sigma^c \theta\, \overbar{F}_{abc}-\frac{i}{24} \varepsilon\sigma_{abcde}\theta\, \overbar{F}^{cde}+\frac{1}{24} \theta\sigma^{cde}\overbar{\varepsilon} \, G_{abcde}+\mbox{c.c.}\ .
\ee
As a check, we take the 
flat space zero flux limit 
\bb
\delta_\varepsilon\theta=\varepsilon\ \ \ \ ,\ \ \ 
\delta_\varepsilon\overbar{\theta}=\overbar{\varepsilon}\ \ \ ,\ \ \ 
\delta x^a=\frac{i}{2} \overbar{\varepsilon}\sigma^a \theta+\frac{i}{2} \varepsilon \sigma^a \overbar{\theta}\ .
\ee
\bb
\delta_L \theta=0\ \ \ ,\ \ \ \delta_L \overbar{\theta}=0\ \ \ ,\ \ \ \delta_L x^a=0\ .
\ee
to arrive at more familiar looking expressions. Note in particular that, for backgrounds without RR and NSNS
fluxes, the Lorentz rotation piece is zero.

\subsection{Kappa symmetry}

The kappa symmetry~\pref{kappa} of~\pref{action} is assured if we choose~\cite{GHMNT}
\bb
\kappa^a=0\ ;
\ee
\bb
\kappa^\alpha=V_i^a \lk(\sigma_a\re)^{\alpha\beta} \lk(h^{ij} {\eta}_{j\beta}
-\frac{\varepsilon^{ij}}{\sqrt{-h}}\overbar{\eta}_{j\beta}\re)\equiv V_i^a \lk(\sigma_a\re)^{\alpha\beta} k^i_\beta\ .
\ee
\bb
\overbar{\kappa}^\alpha=\kappa^\overbar{\alpha}=\overline{\kappa^\alpha}\ .
\ee
Note that the two spinors $\eta_i$ - with $i=0,1$ - have opposite chirality to that of $\theta$. It is more
convenient to rewrite this tranformation using
\bb
k^0=\eta_0-\overbar{\eta}_1\ \ \ ,\ \ \ k^1=-\eta_1+\overbar{\eta}_0\Rightarrow \overbar{k}^0=k^1\ .
\ee
Hence, 
\bb
\kappa^\alpha_{L,R}\equiv\kappa^\alpha\pm\overbar{\kappa}^\alpha=\del_\pm z^M E_M^a \lk(\sigma_a k_{L,R}\re)^\alpha\ ,
\ee
with $k_{L,R}\equiv k^0\pm\overbar{k}^0$ and $\del_\pm\equiv\del_0\pm\del_1$. We will need the kappa transformation in Section~5 to devise a supersymmetry transformation that preserves the light-cone gauge $\sigma^+\theta=0$.

\section{Preserving the light-cone gauge}

In this section, we will formulate the supersymmetry transformations that preserve the light-cone gauge~\pref{lcgauge}. For fixed $\varepsilon^\alpha$ (and hence fixed $\delta_L$), we want to
choose $\delta_\kappa$ such that
\bb
\sigma^+ \lk(\delta_{\kappa+\varepsilon+L} \theta\re)=0\ .
\ee
First, from~\pref{s1} and~\pref{q3}, we know that
\bb\label{xplus}
\delta_{\varepsilon+L} E^+=\delta x^m e_m^+=0\ .
\ee
We next note that Lorentz rotation $\delta_L$ would always preserve the light-cone gauge, 
provided that $\delta_{\varepsilon+\kappa}$ is arranged to do so
\bb
\sigma^+\delta_{\varepsilon+\kappa}\theta=0\ .
\ee
To see this, apply $\sigma^+$ to 
\bb
\delta_L \theta^\alpha=\theta^\beta L_\beta^\alpha
\ee
it is straightforward to check using~\pref{rot1} and~\pref{rot2} that we indeed have
\bb
\sigma^+\delta_L \theta=0\ .
\ee

Hence, we focus on $\delta_{\varepsilon+\kappa}$. From equation~\pref{s4} and \pref{q1}, it is easy to see that\footnote{One needs a `-' index in~\pref{c2}; but we have~\pref{xplus}.}
\bb
\sigma^+\delta_\varepsilon\theta=0\Rightarrow\sigma^+\varepsilon=0\ .
\ee
Therefore, if we look at $\varepsilon$ translations satisfying $\sigma^+\varepsilon=0$, we need
not use kappa transformations to preserve the light-cone gauge. These sixteen supersymmetries ($\varepsilon$ is complex)
would then transform the worldsheet as in
\bb\label{trivial}
\delta_S E^i=\delta_L E^i\ \ \ ,\ \ \ \delta_S E^\alpha=\varepsilon^\alpha+\delta_L E^\alpha\ \ \ ,\ \ \ 
\delta_S E^\overbar{\alpha}=\overline{\delta_S E^\alpha}\ ,
\ee
where $\delta_S$ stands for a supersymmetry transformation preserving the light-cone gauge.
But if $\sigma^+\varepsilon=0$, we also have\footnote{See~\pref{Leq}, \pref{rot1} and~\pref{rot2}, making use
of $\sigma^+\varepsilon=\sigma^+\theta=0$.}
\bb
\delta_L E^i=\delta_L E^\alpha=\delta_L E^\overbar{\alpha}=0\ .
\ee
Equation~\pref{trivial} then entails no interesting structure that may be left unbroken by
a non-trivial worldsheet vacuum. A
BPS condition $\delta_S E^A=0$ implies $\varepsilon=0$. 

Next, we focus on 
\bb
\sigma^+\varepsilon\neq 0
\ee
with $\sigma^-\varepsilon=0$ (or $\varepsilon=\sigma^-\sigma^+\varepsilon$).
The transformation $\delta_\varepsilon$ then takes us out of the light-cone gauge. We need
to use an appropriately tuned kappa transformation to bring us back. 
Choose 
\bb
\sigma^+ k_{L,R}=0\ .
\ee
So that we have
\bb\label{solve}
\sigma^+\varepsilon_{L,R}+\sigma^+ \kappa_{L,R}=0\ ,
\ee
with
\bb\label{with}
\delta_{\kappa+\varepsilon}E^\alpha\pm
\delta_{\kappa+\varepsilon}E^\overbar{\alpha}=\varepsilon_{L,R}^\alpha
+\kappa_{L,R}^\alpha\ .
\ee
We then can write
\bb
\kappa^\alpha_{L,R}=\del_\pm x^m E_m^a \lk(\sigma_a k_{L,R}\re)^\alpha=-\del_\pm x^m e_m^i \lk(\sigma^i k_{L,R}\re)^\alpha+2 p^+ e^+ \lk(\sigma^- k_{L,R}\re)^\alpha\ .
\ee
where we used $x^+=p^+\tau$, and $e^+$ stands for the vielbein component in the `+' direction for
our diagonal metric. It is now straightforward to solve~\pref{solve} for $k_{L,R}$ in terms of $\varepsilon_{L,R}$
\bb
2p^+ e^+ \lk(\sigma^+\sigma^- k_{L,R}\re)=-\sigma^+ \varepsilon_{L,R}\ .
\ee

Putting things together - while focusing on the piece surviving the action of $\sigma^-$ (\ie\ insert
$1=\sigma^+\sigma^-+\sigma^-\sigma^+$ in~\pref{with}) - we get
\bb
\delta_{\kappa+\varepsilon} E^\alpha\pm\delta_{\kappa+\varepsilon}E^\overbar{\alpha}=
\frac{1}{2p^+}\del_\pm x^m \frac{e_m^i}{e^+} \lk(\sigma^i\sigma^+\varepsilon_{L,R}\re)^\alpha\ ;
\ee
\bb
\delta_{\kappa+\varepsilon} E^-=i\overbar{\varepsilon}\sigma^- \theta+i\varepsilon\sigma^- \overbar{\theta}=0\ ;
\ee
\bb
\delta_{\kappa+\varepsilon} E^+=0\ ;
\ee
\bb
\delta_{\kappa+\varepsilon} E^i=i\overbar{\varepsilon}\sigma^i \theta+i\varepsilon\sigma^i \overbar{\theta}\ .
\ee
And more interestingly, including the Lorentz rotation, we find that the supersymmetry
transformation of the worldsheet degrees of freedom in backgrounds with RR and NSNS 
fluxes takes the form
\bb\label{susy1}
\delta_{S} E^\alpha\pm\delta_{S}E^\overbar{\alpha}=\frac{1}{2p^+}\del_\pm x^m \frac{e_m^i}{e^+} \lk(\sigma^i\sigma^+\varepsilon_{L,R}\re)^\alpha+\theta^\beta L_\beta^\alpha\mp\theta^\overbar{\beta} L_\overbar{\beta}^\overbar{\alpha}\ .
\ee
\bb\label{susy2}
\delta_{S} E^-=-\frac{i}{2} \lk(\overbar{\theta}\sigma^-\re)_\alpha \theta^\beta L_\beta^\alpha+\frac{i}{2} \lk(\theta \sigma^-\re)_\overbar{\alpha} \theta^\overbar{\beta} L_\overbar{\beta}^\overbar{\alpha}\ .
\ee
\bb\label{susy3}
\delta_{S} E^+=0\ .
\ee
\bb\label{susy4}
\delta_{S} E^i=i\overbar{\varepsilon}\sigma^i \theta+i\varepsilon\sigma^i \overbar{\theta}\ .
\ee
Hence, these are the remaining sixteen (again $\varepsilon$ is complex) supersymmetry transformations.
We are now ready to look for interesting supersymmetric worldsheet vacua.

\section{Non-trivial supersymmetric vacua}

Equations~\pref{susy1}-\pref{susy4} may be used to look for solutions of the worldsheet fields that are invariant
under supersymmetry transformations
\bb
\delta_S E^A=0\ .
\ee
We also need to make sure that the same supersymmetries that are left unbroken by this statement are
also left unbroken by the background field configuration.
To clarify the structural forms of these equations, we rewrite the complex spinor $\theta$
in terms of two real chiral spinors
\bb
\theta=\theta_1+i\theta_2\ \ \ ,\ \ \ \overbar{\theta}=\theta_1-i\theta_2\ .
\ee
Similarly, we would write $\varepsilon_L=2 \varepsilon_1$, $\varepsilon_R=2 i \varepsilon_2$. 
Equation~\pref{susy4} then becomes
\bb\label{w1}
\varepsilon_1\sigma^i \theta_1+\varepsilon_2 \sigma^i \theta_2=0\ .
\ee
While equation~\pref{susy2} is
\bb\label{w2}
\lk( \theta_1 \sigma^{-ab}\theta_1+\theta_2 \sigma^{-ab}\theta_2\re) L_{ab}=0\ .
\ee
And finally equation~\pref{susy1} is
\bb\label{w3}
\frac{\del_\pm x^m}{p^+} \frac{e^i_m}{e^+} \lk( \sigma^i \sigma^+\varepsilon_{1,2}\re)
-\frac{1}{2} L_{ab} \lk(\sigma^{ab} \theta_{1,2} \re)=0\ .
\ee
With
\bb
\sigma^+ \varepsilon_{1,2}\neq 0\ \ \ ,\ \ \ \sigma^+ \theta_{1,2}=0\ ,
\ee
and
\bbb\label{Lab}
L_{ab}&=&-\frac{3}{2} F_{abc}^I \lk(\varepsilon_1 \sigma^c \theta_1-\varepsilon_2\sigma^c\theta_2\re)
-\frac{1}{12} F^{I\,cde} \lk(\varepsilon_1 \sigma_{abcde}\theta_1-\varepsilon_2 \sigma_{abcde}\theta_2\re) \nonumber \\
&+&\frac{3}{2} F^R_{abc} \lk(\varepsilon_2 \sigma^c\theta_1+\varepsilon_1 \sigma^c \theta_2\re)
+\frac{1}{12} F^{R\,cde}\lk(\varepsilon_2 \sigma_{abcde}\theta_1+\varepsilon_1\sigma_{abcde}\theta_2\re) \nonumber \\
&+&\frac{1}{12} G_{abcde} \lk(\varepsilon_1 \sigma^{cde}\theta_1+\varepsilon_2 \sigma^{cde} \theta_2\re)
\ ,
\eee
equations~\pref{w1}, \pref{w2} and~\pref{w3} are our  BPS conditions. We note that $F^R_{abc}$ is roughly NSNS flux; and $F^I_{abc}$ contains RR flux\footnote{As a quick check of the conditions we are exploring, we note that the flat space zero flux limit with a unidirectional wave corresponds to the
requirement $\del_+ x^a=0$ or $\del_-x=0$; \ie\ as expected, exclusively left or right moving excitations on the string are BPS with eight supersymmetries since~\pref{susy1} dictates that either $\varepsilon_R$ or $\varepsilon_L$ must vanish.}.

\section{A supersymmetric solution}

Let us start by focusing on equation~\pref{w1}. We may satisfy this statement if we arrange
\bb\label{D1}
\varepsilon_1=\pm\varepsilon_2\ \ \ ,\ \ \ \theta_1=\mp\theta_2\ .
\ee
From~\pref{Lab}, we see that this case corresponds to a Lorentz rotation involving RR fluxes (both $\chi$ and $H^{(2)}$) and no NSNS fluxes. 
In contrast, the alternative statement
\bb
\varepsilon_1=\pm\varepsilon_2\ \ \ ,\ \ \ \theta_1=\pm\theta_2
\ee
leads to 
no RR couplings in~\pref{Lab}; and condition~\pref{w1} needs to be imposed
separately. The latter is a stringent statement that would imply $\theta_{1,2}=0$\footnote{One may not set $\varepsilon_1$ equal to $\theta_1$ since $\sigma^+\varepsilon_1\neq 0$.}. Hence, the interesting case is given by~\pref{D1}, when RR fluxes are present. We then consider this scenario, and analyze the
remaining equations~\pref{w2} and~\pref{w3}. This leaves up with eight unbroken supersymmetries.

First, note that for RR fluxes, we have $L_{-i}=L_{+i}=0$, $L_{ij}\neq 0$, $L_{+-}\neq 0$. Our goal is 
to show that it is now possible to have a non-trivial vev for the spinors $\theta_{1,2}$. To simplify matters, let us consider an electric 
field strength $F^{I\, -+i}\neq 0$, $F^{I\, ijk}=0$. Perhaps our closed string is
in the vicinity of D1 strings. We then have
\bb
L_{ij}^{el}= 2 F^{I\, -+k} \varepsilon_1 \sigma^+\sigma_{ijk} \sigma^- \theta_1\ ;
\ee
\bb
L_{-+}^{el}=-12 F^{I\, -+i}\varepsilon_1 \sigma^+ \sigma_i \sigma^- \theta_1\ .
\ee
We look for the simplest configuration for the bosonic fields
\footnote{In the case where one may wrap the closed string along a compact cycle of the geometry, it was shown in~\cite{VVSlargeM} that it may be possible to cancel (semiclassically) the effect
of RR fluxes in certain situations. Here, we consider a more generic scenario.}
\bb
\del_\tau{x}^m=0\ \ \ ,\ \ \ \del_\sigma {x^m}=0\ .
\ee
From~\pref{w2}, one gets
\bb\label{z1}
\lk(\theta_1 \sigma^{-ij}\theta_1\re)\lk(\varepsilon_1 \sigma^+ \sigma_{ijk} \sigma^- \theta_1\re) F^{I\,-+k}=0\ .
\ee
And from~\pref{w3}, one has
\bb\label{z2}
F^{I\,-+k}\lk(\varepsilon_1 \sigma^+ \sigma_{ijk} \sigma^- \theta_1\re) \lk(\sigma^- \sigma^{ij}\theta_1\re)
+6 F^{I\,-+k} \lk(\varepsilon_1 \sigma^+ \sigma_k \sigma^-\theta_1\re) \lk(\sigma^-\theta_1\re)=0\ .
\ee
We now need to solve~\pref{z1} and~\pref{z2} for $\theta_1$ for fixed non-zero $\varepsilon_1$.
The problem may be recast into a system of linear equations for a number of bosonic
variable $n_i$ and $m_{ijk}=m_{[ijk]}$ by writing the most general expression
\bb\label{sol}
\varepsilon_1=n_k \sigma^k \sigma^- \theta_1+m_{klm} \sigma^{klm} \sigma^- \theta_1\ .
\ee
The implication is that we shall invert this equation at the end for $\theta_1$.
Substitute~\pref{sol} in~\pref{z1} and~\pref{z2} and use Fierz identities to rearrange things (see Appendix A). To write the result in a suggestive form, 
let us define the objects
\bb
A^{ij}\equiv \theta_1 \sigma^{-ij}\theta_1\ \ \ ,\ \ \ A_0^{ij}\equiv  \sigma^{-ij}\theta_1\ \ \ ,\ \ \ A_0\equiv  \sigma^-\theta_1\ .
\ee
Equation~\pref{z1} then becomes
\bb\label{n1}
A^{ij} \GG_{ij,kl} A^{kl}=0\ .
\ee
where the `metric' is written as
\bb\label{Gijkl}
\GG_{ij,kl}=-\eta_{ik}\eta_{jl} F^{I\,-+m} n_m
+\eta_{jl} F^{I\,-+}_{\ \ \ \ \ k} n_i +\eta_{jl}F^{I\,-+}_{\ \ \ \ \ i} n_k
-3 F^{I\,-+}_{\ \ \ \ \ k} m_{ijl}-3 F^{I\,-+}_{\ \ \ \ \ i} m_{klj}\ .
\ee
Notice that $\GG_{ij,kl}=\GG_{kl,ij}$; and that we may antisymmetrize $[ij]$ and $[kl]$ if desired.
Equation~\pref{z2} involves slightly more work. One gets an expression looking like
\bb
A^{ij} {\GG'}_{ij,kl} A^{kl}_0+{\HH'}_{ij} A^{ij} A_0=0
\ee
where $\GG'$ includes $\GG$ (the part symmetric in the indices $(ij,kl)$), as well as additional
pieces in the antisymmetric combination of the indices. While there is no obvious symmetric
structure in the product $A^{ij} A_0^{kl}$ in exchanging $(ij)$ with $(kl)$, it is possible to
unravel the antisymmetric part and combine it with what we have labeled $\HH'_{ij}$. To do this,
one uses Fierz rearrangements, such as
\bbb
\lk(\theta \sigma^{-ij}\theta\re) \lk(\sigma^{-kl}\theta\re)&=& 
-\frac{1}{16^2}\lk(\theta \sigma^{-mn}\theta\re) \nonumber \\
& &\lk[
\mbox{Tr}\lk(\sigma^{ij}\sigma_{mn} \sigma^{kl}\re)\lk(\sigma^- \theta\re)
+\frac{1}{2} \mbox{Tr} \lk(\sigma^{ij}\sigma_{mn}\sigma^{kl}\sigma_{pq}\re) \lk(
\sigma^{-pq}\theta\re)\re ]\ .
\eee
One then can rewrite
\bb
A^{ij} {\GG'}_{ij,kl} A^{kl}_0=A^{ij} {\GG}_{ij,kl} A^{kl}_0+A^{ij} \AA^{kl,mn}_{ij}{\GG'}^{A}_{kl,mn} A_0\ ,
\ee
where $\AA^{kl,mn}_{ij}$ may be computed from the traces of gamma matrices, and ${\GG'}^{A}_{kl,mn}$ stands for the part of $\GG'$ antisymmetric under exchange of $(kl)$ with $(mn)$. 
When the dust settles, one gets equation~\pref{z2} recast into the form
\bb\label{n2}
A^{ij} {\GG}_{ij,kl} A^{kl}_0+{\HH}_{ij} A^{ij} A_0=0
\ee
where ${\GG}_{ij,kl}$ is as in~\pref{Gijkl}; and
\bb
\HH_{ij}=\frac{51}{2} F^{I\, -+k} m_{kij}+\frac{9}{2} F^{I\, -+}_{\ \ \ \ [i} n_{j]}\ .
\ee

To solve for $n_i$ and $m_{ijk}$ in~\pref{n1} and~\pref{n2}, one may use a Clifford algebra basis such that
\bb
A^{23}=A^{45}=A^{67}=A^{89}=0
\ee
since the spinors obey Fermi statistics and $\sigma^-$ is a symmetric matrix. One then needs 
\bb\label{f1}
\GG_{ij,kl}=0\ \ \ \mbox{if $(ij) \mbox{ or } (kl)\neq \lk\{23,45,67,89\re\}$}\ .
\ee
And
\bb\label{f2}
{\HH}_{ij} A^{ij}=0\Rightarrow {\HH}_{ij}=0\ \ \ \mbox{if $(ij)\neq \lk\{23,45,67,89\re\}$}\ .
\ee

Let us now specialize to a simple background field configuration to be more explicit. Arrange a network of parallel D1 branes stretched
along the 1 direction, filling all of the space directions $i=2\ldots 8$; we would then have a constant flux pointing in
the 9 direction
\bb
F^{I\,-+9}=\mbox{constant}\ \ \ ,\ \ \ \mbox{All other components zero.}
\ee
Notice also that this background leaves unbroken the supersymmetries given by~\pref{D1}. 
Equations~\pref{f1} and~\pref{f2} are then satisfied if
\bb
n^i=0\ \ ,\ \ \mbox{and}\ \ \ m^{ijk}=0\ \ \ \mbox{except if $(ijk)=\lk\{923,945,967\re\}$}\ .
\ee
The remaining three $m_{ijk}$'s are otherwise arbitrary. We now also see 
that we may invert~\pref{sol} to
solve for a non-trivial vev for $\theta_1$ for {\em any} of the eight supersymmetries $\varepsilon_1$. We have hence shown that the vacuum of the worldsheet can get polarized in the presence of RR fluxes in directions correlating with the orientation of the D1 branes. Note also that we have three free
bosonic degrees of freedom, one for each of the planes $(23)$, $(45)$ and $(67)$.

\section{Discussion}

We have shown that the presence of background RR fluxes can polarizes worldsheet spinor degrees of
freedom. We contrasted this with the case involving NSNS fluxes where no polarization occurs. We expect this phenomenon to be a general one. Furthermore, with non-trivial vacuum degeneracy, we
can speculate that the worldsheet theory may admit worldsheet `soliton' configurations - with interesting spinor profiles - that can source RR moments (we would still expect zero RR monopole moment). In particular, if one turns on a profile for the bosonic excitations $x^m(\tau,\sigma)$, one may locally lift some of the flat directions of our solution of the previous section towards zero vev (see equation~\pref{w3}). It would then be interesting to see if it is possible to lock the spinors in one polarization state, say $923$, in some limiting regime on the worldsheet; and then lock them in another state, say $945$, in another limiting regime. Such configurations would source RR flux, yet they would be built from closed string degrees of freedom. It would be hoped that the identification of such configurations can shed light on the interplay between open and closed string dynamics.

Beyond looking for such worldsheet solitons, other interesting directions involve an analysis of the structure of the degeneracy in the scenario we presented. In particular, using~\cite{VVSRR}, we may look at the dynamics of small fluctuations about 
spinor vevs. It would also be interesting to consider other toy systems that can help one develop
intuition about the problem. In particular, an interesting scenario is one involving a profile
of the RR axion (D(-1) brane charge), along with NSNS 3-form flux. As we see from~\pref{chi}, such a configuration may yield vevs for the spinors as well. The instantonic nature of the source of the RR field
in such a scenario may help clarify the issue of encoding of D-brane worldvolume directions onto a soliton in
1+1 dimensions. We hope to report on some of these issues in the future.

\section{Appendices}
\subsection{Appendix A: Spinor conventions}

In this work, we are using the Clifford algebra convention used in~\cite{HOWEWEST, GHMNT,VVSRR}. The $16\times 16$ gamma matrices satisfy
\bb
\lk\{\sigma^a,\sigma^b\re\}=2\eta^{ab}\ ,
\ee
where $\eta^{ab}$ is the flat metric with signature $+---\cdots$. We then have
$\sigma^+\sigma^-+\sigma^-\sigma^+=1$. The $\sigma^a$'s are real $\overline{\sigma^a}=\sigma^a$;
$\sigma^a$, $\sigma^{abcd}$, and $\sigma^{abcde}$ are symmetric; while $\sigma^{ab}$ and $\sigma^{abc}$ are antisymmetric. We also note the useful rearrangement
\bb
Q_{\alpha\beta}=\frac{1}{16} \lk(
\mbox{Tr}\lk[Q\sigma_a\re] \sigma^a_{\alpha\beta}
-\frac{1}{3!}\mbox{Tr}\lk[Q\sigma_{abc}\re] \sigma^{abc}_{\alpha\beta}
+\frac{1}{5!}\mbox{Tr}\lk[Q\sigma_{abcde}\re] \sigma^{abcde}_{\alpha\beta}\re)\ .
\ee
And $\sigma^{abcde}$ is self-dual. 

\subsection{Appendix B: Normal coordinate expansion in superspace}

The normal coordinate expansion technique in superspace allows one to unravel the component form of superspace expressions. In this work, we used it to write the explicit form of the variation $\delta E^A$. Using~\cite{GRISARU} and~\cite{ATICKDHAR}, one writes
\bb
{E'}^A=E^A+\delta E^A+\frac{1}{2!} \delta^2 E^A+\cdots
\ee
Each variation, evaluated at zeroth order, may be computed using the relations
\bb
\delta E^A=Dy^A+y^CE^V T_{BC}^{\ \ \ \ A}\ ;
\ee
\bb
\delta Dy^A=-y^B E^C y^D R_{DCB}^{\ \ \ \ \ A}\ .
\ee
Here, $y^A$ is the displacement in normal coordinates from the point $z^M=0$ for $M$ fermionic. The supertorsion and superriemann tensors may be found in~\cite{HOWEWEST} in the same notation used in this work.

\providecommand{\href}[2]{#2}\begingroup\raggedright\endgroup

\end{document}

%% file: macros.tex

\newcommand{\bb}{\begin{equation}}
\newcommand{\ee}{\end{equation}}
\newcommand{\bbb}{\begin{eqnarray}}
\newcommand{\eee}{\end{eqnarray}}
\newcommand{\diag}{\mbox{diag }}
\newcommand{\Str}{\mbox{STr }}
\newcommand{\Tr}{\mbox{Tr }}
\newcommand{\Det}{\mbox{Det }}
\newcommand{\C}[2]{{\lk [{#1},{#2}\re ]}}
\newcommand{\AC}[2]{{\lk \{{#1},{#2}\re \}}}
\newcommand{\kk}{\hspace{.5em}}
\newcommand{\vc}[1]{\mbox{$\vec{{\bf #1}}$}}
\newcommand{\mc}[1]{\mathcal{#1}}
\newcommand{\del}{\partial}
\newcommand{\lk}{\left}
\newcommand{\ave}[1]{\mbox{$\langle{#1}\rangle$}}
\newcommand{\re}{\right}
\newcommand{\pd}[1]{\frac{\del}{\del #1}}
\newcommand{\pdd}[2]{\frac{\del^2}{\del #1 \del #2}}
\newcommand{\Dd}[1]{\frac{d}{d #1}}
\newcommand{\sech}{\mbox{sech}}
\newcommand{\pref}[1]{(\ref{#1})}

\newcommand
{\sect}[1]{\vspace{20pt}{\LARGE}\noindent
{\bf #1:}}
\newcommand
{\subsect}[1]{\vspace{20pt}\hspace*{10pt}{\Large{$\bullet$}}\mbox{ }
{\bf #1}}
\newcommand
{\subsubsect}[1]{\hspace*{20pt}{\large{$\bullet$}}\mbox{ }
{\bf #1}}

\def\ie{{\it i.e.}}
\def\eg{{\it e.g.}}
\def\cf{{\it c.f.}}
\def\etal{{\it et.al.}}
\def\etc{{\it etc.}}

\def\e{{\mbox{{\bf e}}}}
\def\AA{{\cal A}}
\def\BB{{\cal B}}
\def\CC{{\cal C}}
\def\DD{{\cal D}}
\def\EE{{\cal E}}
\def\FF{{\cal F}}
\def\GG{{\cal G}}
\def\HH{{\cal H}}
\def\II{{\cal I}}
\def\JJ{{\cal J}}
\def\KK{{\cal K}}
\def\LL{{\cal L}}
\def\MM{{\cal M}}
\def\NN{{\cal N}}
\def\OO{{\cal O}}
\def\PP{{\cal P}}
\def\QQ{{\cal Q}}
\def\RR{{\cal R}}
\def\SS{{\cal S}}
\def\TT{{\cal T}}
\def\UU{{\cal U}}
\def\VV{{\cal V}}
\def\WW{{\cal W}}
\def\XX{{\cal X}}
\def\YY{{\cal Y}}
\def\ZZ{{\cal Z}}

\def\sinh{{\rm sinh}}
\def\cosh{{\rm cosh}}
\def\tanh{{\rm tanh}}
\def\sgn{{\rm sgn}}
\def\det{{\rm det}}
\def\trace{{\rm Tr}}
\def\exp{{\rm exp}}
\def\sh{{\rm sh}}
\def\ch{{\rm ch}}

\def\ell{{\it l}}
\def\str{{\it str}}
\def\lp{\ell_{{\rm pl}}}
\def\blp{\overline{\ell}_{{\rm pl}}}
\def\ls{\ell_{{\str}}}
\def\bls{{\bar\ell}_{{\str}}}
\def\bM{{\overline{\rm M}}}
\def\gs{g_\str}
\def\gym{{g_{Y}}}
\def\geff{g_{\rm eff}}
\def\eff{{\rm eff}}
\def\r11{R_{11}}
\def\kel{{2\kappa_{11}^2}}
\def\kten{{2\kappa_{10}^2}}
\def\lpten{{\lp^{(10)}}}
\def\alp{{\alpha '}}
\def\alpe{{{\alpha_e}}}
\def\le{{{l}_e}}
\def\aleff{{\alp_{eff}}}
\def\sqaleff{{\alp_{eff}^2}}
\def\tgs{{\tilde{g}_s}}
\def\talp{{{\tilde{\alpha}}'}}
\def\tlp{{\tilde{\ell}_{{\rm pl}}}}
\def\tr11{{\tilde{R}_{11}}}
\def\wtilde{\widetilde}
\def\what{\widehat}
\def\hlp{{\hat{\ell}_{{\rm pl}}}}
\def\hr11{{\hat{R}_{11}}}
\def\hf{{\textstyle\frac12}}
\def\coeff#1#2{{\textstyle{#1\over#2}}}
\def\CY{Calabi-Yau}
\def\lessapprox{\;{\buildrel{<}\over{\scriptstyle\sim}}\;}
\def\greaterapprox{\;{\buildrel{>}\over{\scriptstyle\sim}}\;}
\def\inbar{\,\vrule height1.5ex width.4pt depth0pt}
\def\IC{\relax\hbox{$\inbar\kern-.3em{\rm C}$}}
\def\IR{\relax{\rm I\kern-.18em R}}
\def\IP{\relax{\rm I\kern-.18em P}}
\def\Z{{\bf Z}}
\def\R{{\bf R}}
\def\One{{1\hskip -3pt {\rm l}}}
\def\sst{\scriptscriptstyle}
\def\osc{{\rm\sst osc}}
\def\lam{\lambda}
\def\lc{{\sst LC}}
\def\pr{{\sst \rm pr}}
\def\cl{{\sst \rm cl}}
\def\D{{\sst D}}
\def\bh{{\sst BH}}
\def\vev#1{\langle#1\rangle}

%% file: bps.bbl
\begin{thebibliography}{10}

\bibitem{MALDA1}
J.~Maldacena, ``The large N limit of superconformal field theories and
  supergravity,'' \href{http://xxx.lanl.gov/abs/hep-th/9711200}{{\tt
  hep-th/9711200}}.

\bibitem{KLEB}
S.~S. Gubser, I.~R. Klebanov, and A.~M. Polyakov, ``Gauge theory correlators
  from noncritical string theory,'' {\em Phys. Lett.} {\bf B428} (1998) 105,
  \href{http://xxx.lanl.gov/abs/hep-th/9802109}{{\tt hep-th/9802109}}.

\bibitem{WITHOLO}
E.~Witten, ``Anti-de Sitter space and holography,''
  \href{http://xxx.lanl.gov/abs/hep-th/9802150}{{\tt hep-th/9802150}}.

\bibitem{BVW}
N.~Berkovits, C.~Vafa, and E.~Witten, ``Conformal field theory of AdS
  background with Ramond-Ramond flux,'' {\em JHEP} {\bf 03} (1999) 018,
  \href{http://xxx.lanl.gov/abs/hep-th/9902098}{{\tt hep-th/9902098}}.

\bibitem{RAHMRAJ}
J.~Rahmfeld and A.~Rajaraman,
``The GS string action on AdS(3) x S(3) with Ramond-Ramond charge,''
Phys.\ Rev.\ D {\bf 60}, 064014 (1999)

\bibitem{METSAEV}
R.~R.~Metsaev and A.~A.~Tseytlin,
``Type Iib Green-Schwarz Superstrings In Ads(5) X S(5) From The Supercoset Approach,''
J.\ Exp.\ Theor.\ Phys.\  {\bf 91}, 1098 (2000)

\bibitem{PESANDO}
I.~Pesando, ``On the fixing of the kappa gauge symmetry on AdS and flat
  background: The lightcone action for the type IIB string on AdS(5) x S(5),''
  {\em Phys. Lett.} {\bf B485} (2000) 246--254,
  \href{http://xxx.lanl.gov/abs/http://arXiv.org/abs/hep-th/9912284}{{\tt
  http://arXiv.org/abs/hep-th/9912284}}.

\bibitem{VVSlargeM}
V.~Sahakian, ``Holography with Ramond-Ramond fluxes,'' JHEP 0212 (2002) 030
  \href{http://xxx.lanl.gov/abs/http://arXiv.org/abs/hep-th/0209179}{{\tt
  http://arXiv.org/abs/hep-th/0209179}}.

\bibitem{VVSRR}
V.~Sahakian, ``Closed strings in Ramond-Ramond backgrounds,'' JHEP 0404 (2004) 026
  \href{http://xxx.lanl.gov/abs/http://arXiv.org/abs/hep-th/0402037}{{\tt
  http://arXiv.org/abs/hep-th/0402037}}.

\bibitem{MIZO}
Shun'ya Mizoguchi, Takeshi Mogami, Yuji Satoh
, ``Penrose limits and Green-Schwarz strings,'' Class.Quant.Grav. 20 (2003) 1489-1502
  \href{http://xxx.lanl.gov/abs/http://arXiv.org/abs/hep-th/0209043}{{\tt
  http://arXiv.org/abs/hep-th/0209043}}.

\bibitem{MAROLF}
Donald Marolf, Luca Martucci, Pedro J. Silva
, ``The explicit form of the effective action for F1 and D-branes,'' Class. Quantum Grav. 21 (21 May 2004) S1385-S1390
  \href{http://xxx.lanl.gov/abs/http://arXiv.org/abs/hep-th/0404197}{{\tt
  http://arXiv.org/abs/hep-th/0404197}}.

\bibitem{PEDRO}
Luca Martucci, Pedro J. Silva, ``On type II superstrings in bosonic backgrounds: the role of fermions and  T-duality,'' JHEP 0304 (2003) 004
  \href{http://xxx.lanl.gov/abs/http://arXiv.org/abs/hep-th/0303102}{{\tt
  http://arXiv.org/abs/hep-th/0303102}}.

\bibitem{RUBAKOV}
V.~Rubakov, ``Classical theory of gauge fields,'' Princeton University Press (2002)

\bibitem{GHMNT}
M.~T. Grisaru, P.~Howe, L.~Mezincescu, B.~Nilsson, and P.~K. Townsend, ``N=2
  superstrings in a supergravity background,'' {\em Phys. Lett.} {\bf B162}
  (1985) 116.

\bibitem{HOWEWEST}
P.~S. Howe and P.~C. West, ``The complete N=2, d = 10 supergravity,'' {\em
  Nucl. Phys.} {\bf B238} (1984) 181.

\bibitem{GRISARU}
M.~T. Grisaru, M.~E. Knutt-Wehlau, and W.~Siegel, ``A superspace normal
  coordinate derivation of the density formula,'' {\em Nucl. Phys.} {\bf B523}
  (1998) 663--679,
  \href{http://xxx.lanl.gov/abs/http://arXiv.org/abs/hep-th/9711120}{{\tt
  http://arXiv.org/abs/hep-th/9711120}}.

\bibitem{ATICKDHAR}
J.~J. Atick and A.~Dhar, ``Normal coordinates, theta expansion and strings on
  curved superspace,'' {\em Nucl. Phys.} {\bf B284} (1987) 131.

\end{thebibliography}
